\documentclass[aps,prl,twocolumn,superscriptaddress,showpacs,preprintnumbers,amsmath,amssymb]{revtex4}

\usepackage{graphicx} % Include figure files
\usepackage{dcolumn}  % Align table columns on decimal point
\usepackage{subfigure}

\newcommand{\sqs} {\ensuremath{ \sqrt{s} }}
\newcommand{\RM}  {\ensuremath{ M_{\mathrm{rec}} }}
\newcommand{\RMF} {\ensuremath{ M^\mathrm{fit}_{\mathrm{rec}} }}

\newcommand{\dd}    {\ensuremath{ D^{(*)^{\pm}}{D}{}^{*\mp}}}
\newcommand{\ddch}  {\ensuremath{ D^{(*)+} D^{*-}        }}
\newcommand{\dpdm}  {\ensuremath{ D^{*+} D^{*-}          }}
\newcommand{\dpdstm}{\ensuremath{ D^+ D^{*-}             }}

\newcommand{\gisr}{\ensuremath{\gamma_{\mathrm{isr}}}}
\newcommand{\gevc}{\ensuremath{\, {\mathrm{GeV}/c^2} }}
\newcommand{\mevc}{\ensuremath{\, {\mathrm{MeV}/c^2} }}
\newcommand{\gev} {\ensuremath{\, {\mathrm{GeV}} }}
\newcommand{\mev} {\ensuremath{\, {\mathrm{MeV}} }}
\newcommand{\ee}  {\ensuremath{ e^+ e^- }}

\newcommand{\eedd}     {\ensuremath{ e^+e^- \to D^{(*)\pm}{D}{}^{*\mp}}}
\newcommand{\eeddch}   {\ensuremath{ e^+e^- \to D^{(*)+} D^{*-} }}
\newcommand{\eedpdm}   {\ensuremath{ e^+e^- \to D^{*+} D^{*-}   }}
\newcommand{\eedpdstm} {\ensuremath{ e^+e^- \to D^+ D^{*-}      }}

\newcommand{\eeddchg}  {\ensuremath{ e^+e^- \to D^{(*)+} D^{*-} \gisr}}
\newcommand{\eedpdmg}  {\ensuremath{ e^+e^- \to D^{*+} D^{*-}   \gisr}}
\newcommand{\eedpdstmg}{\ensuremath{ e^+e^- \to D^+    D^{*-}   \gisr}}

\newcommand{\dpdmpi} {\ensuremath{ e^+e^- \to D^{*+} D^{*-}\pi^0_{\mathrm{miss}} \gisr}}
\newcommand{\dpdmis} {\ensuremath{ e^+e^- \to D^{*+} D^{*-}\pi^0}}

\newcommand{\dpdstmis}  {\ensuremath{ e^+e^- \to D^+ D^{*-} \pi^0 }}
\newcommand{\dndstg}    {\ensuremath{ e^+e^- \to D^0 D^{*-}\pi^+_{\mathrm{miss}} \gisr}}

\begin{document}

\title{ \quad\\[0.5cm] Measurement of the near-threshold \eedd\ cross
section using initial-state radiation}

%\input{author}
%%% Paper:    e+ e- -> D(*) D*-bar
%%% Journal:  Physical Review Letters
%%% Contacts: G. Pakhlova (galya@itep.ru)
%%% Non-responding authors or those who said NO are commented out.
%%% ====================================================================
%%% Click the RELOAD button on your web browser to see the updated file.
%%% ====================================================================
%%% Use \input{author} to insert this material into your latex file.
%%%%% Force institutions to appear in alphabetical order when typeset.
\affiliation{Budker Institute of Nuclear Physics, Novosibirsk}
%%%\affiliation{Chiba University, Chiba}
\affiliation{Chonnam National University, Kwangju}
\affiliation{University of Cincinnati, Cincinnati, Ohio 45221}
\affiliation{Department of Physics, Fu Jen Catholic University, Taipei}
%%%\affiliation{Justus-Liebig-Universit\"at Gie\ss{}en, Gie\ss{}en}
\affiliation{The Graduate University for Advanced Studies, Hayama, Japan}
\affiliation{Gyeongsang National University, Chinju}
\affiliation{University of Hawaii, Honolulu, Hawaii 96822}
\affiliation{High Energy Accelerator Research Organization (KEK), Tsukuba}
%%%\affiliation{Hiroshima Institute of Technology, Hiroshima}
\affiliation{University of Illinois at Urbana-Champaign, Urbana, Illinois 61801}
\affiliation{Institute of High Energy Physics, Chinese Academy of Sciences, Beijing}
\affiliation{Institute of High Energy Physics, Vienna}
\affiliation{Institute of High Energy Physics, Protvino}
\affiliation{Institute for Theoretical and Experimental Physics, Moscow}
\affiliation{J. Stefan Institute, Ljubljana}
\affiliation{Kanagawa University, Yokohama}
\affiliation{Korea University, Seoul}
%%%\affiliation{Kyoto University, Kyoto}
\affiliation{Kyungpook National University, Taegu}
\affiliation{Swiss Federal Institute of Technology of Lausanne, EPFL, Lausanne}
\affiliation{University of Ljubljana, Ljubljana}
\affiliation{University of Maribor, Maribor}
\affiliation{University of Melbourne, Victoria}
\affiliation{Nagoya University, Nagoya}
\affiliation{Nara Women's University, Nara}
\affiliation{National Central University, Chung-li}
\affiliation{National United University, Miao Li}
\affiliation{Department of Physics, National Taiwan University, Taipei}
\affiliation{H. Niewodniczanski Institute of Nuclear Physics, Krakow}
\affiliation{Nippon Dental University, Niigata}
\affiliation{Niigata University, Niigata}
\affiliation{University of Nova Gorica, Nova Gorica}
\affiliation{Osaka City University, Osaka}
\affiliation{Osaka University, Osaka}
\affiliation{Panjab University, Chandigarh}
\affiliation{Peking University, Beijing}
%%%\affiliation{University of Pittsburgh, Pittsburgh, Pennsylvania 15260}
%%%\affiliation{Princeton University, Princeton, New Jersey 08544}
\affiliation{RIKEN BNL Research Center, Upton, New York 11973}
%%%\affiliation{Saga University, Saga}
\affiliation{University of Science and Technology of China, Hefei}
\affiliation{Seoul National University, Seoul}
\affiliation{Shinshu University, Nagano}
\affiliation{Sungkyunkwan University, Suwon}
\affiliation{University of Sydney, Sydney NSW}
\affiliation{Tata Institute of Fundamental Research, Bombay}
\affiliation{Toho University, Funabashi}
\affiliation{Tohoku Gakuin University, Tagajo}
\affiliation{Tohoku University, Sendai}
\affiliation{Department of Physics, University of Tokyo, Tokyo}
\affiliation{Tokyo Institute of Technology, Tokyo}
\affiliation{Tokyo Metropolitan University, Tokyo}
\affiliation{Tokyo University of Agriculture and Technology, Tokyo}
%%%\affiliation{Toyama National College of Maritime Technology, Toyama}
%%%\affiliation{University of Tsukuba, Tsukuba}
\affiliation{Virginia Polytechnic Institute and State University, Blacksburg, Virginia 24061}
\affiliation{Yonsei University, Seoul}
  \author{G.~Pakhlova}\affiliation{Institute for Theoretical and Experimental Physics, Moscow} % ITEP
  \author{K.~Abe}\affiliation{High Energy Accelerator Research Organization (KEK), Tsukuba} % KEK
% \author{K.~Abe}\affiliation{Tohoku Gakuin University, Tagajo} % TohokuGakuin
% \author{N.~Abe}\affiliation{Tokyo Institute of Technology, Tokyo} % TIT
  \author{I.~Adachi}\affiliation{High Energy Accelerator Research Organization (KEK), Tsukuba} % KEK
  \author{H.~Aihara}\affiliation{Department of Physics, University of Tokyo, Tokyo} % Tokyo
  \author{D.~Anipko}\affiliation{Budker Institute of Nuclear Physics, Novosibirsk} % BINP
% \author{K.~Aoki}\affiliation{Nagoya University, Nagoya} % Nagoya
% \author{K.~Arinstein}\affiliation{Budker Institute of Nuclear Physics, Novosibirsk} % BINP
% \author{Y.~Asano}\affiliation{University of Tsukuba, Tsukuba} % Tsukuba
% \author{T.~Aso}\affiliation{Toyama National College of Maritime Technology, Toyama} % Toyama
  \author{V.~Aulchenko}\affiliation{Budker Institute of Nuclear Physics, Novosibirsk} % BINP
  \author{T.~Aushev}\affiliation{Swiss Federal Institute of Technology of Lausanne, EPFL, Lausanne}\affiliation{Institute for Theoretical and Experimental Physics, Moscow} % ITEP
% \author{T.~Aziz}\affiliation{Tata Institute of Fundamental Research, Bombay} % Tata
% \author{S.~Bahinipati}\affiliation{University of Cincinnati, Cincinnati, Ohio 45221} % Cincinnati
  \author{A.~M.~Bakich}\affiliation{University of Sydney, Sydney NSW} % Sydney
\author{V.~Balagura}\affiliation{Institute for Theoretical and Experimental Physics, Moscow} % ITEP
% \author{Y.~Ban}\affiliation{Peking University, Beijing} % Peking
% \author{S.~Banerjee}\affiliation{Tata Institute of Fundamental Research, Bombay} % Tata
  \author{E.~Barberio}\affiliation{University of Melbourne, Victoria} % Melbourne
% \author{M.~Barbero}\affiliation{University of Hawaii, Honolulu, Hawaii 96822} % Hawaii
% ...
% Ljubljana
% \author{D.~Z\"urcher}\affiliation{Swiss Federal Institute of Technology of Lausanne, EPFL, Lausanne} % Lausanne
\collaboration{The Belle Collaboration}

\begin{abstract}
We report a measurement of the exclusive \eedd\ cross section as a
function of center-of-mass energy near the \dd\ threshold with initial
state radiation. A partial reconstruction technique is used to
increase the efficiency and to suppress background. The analysis is
based on a data sample collected with the Belle detector at the
$\Upsilon(4S)$ resonance and nearby continuum with an integrated
luminosity of $547.8$ $\mathrm{fb}^{-1}$ at the KEKB asymmetric-energy
\ee\ collider.
\end{abstract}


\maketitle
\setcounter{footnote}{0}

Exclusive \ee\ hadronic cross sections to final states with charm
meson pairs are of special interest because they provide information
on the spectrum of $J^{PC}=1^{--}$ charmonium states above the
open-charm threshold, which is poorly understood~\cite{seth}.  The
observation of the charmonium-like $Y(4260)$ state, seen only in
$J/\psi \pi\pi(KK)$ final states~\cite{babar1,cleo2,cleo3,belle}, has
stimulated renewed interest in this field.  Curiously, the $Y(4260)$
peak position is close to a local minimum of the total hadronic cross
section~\cite{bes1}.  The large branching fraction to $J/\psi\pi\pi$
inferred from the total hadronic cross section is unexpected for a
conventional charmonium state with such a large mass and width.  In a
recent measurement~\cite{dd} the $e^+e^- \to D \overline{D}$ cross
section is described by known charmonium states without a significant
$Y(4260)$ contribution.  A study of the production cross-section of
the charmed meson pairs in this energy range could help clarify this
intriguing situation.

In this paper we report a measurement of exclusive cross sections for
\eedpdm\ and \eedpdstm~\cite{foot1} using initial state radiation
(ISR). The data used for this analysis correspond to an integrated
luminosity of $547.8\,\mathrm{fb}^{-1}$ collected with the Belle
detector~\cite{det} at the $\Upsilon(4S)$ resonance and nearby
continuum at the KEKB asymmetric-energy \ee\ collider~\cite{kekb}.

To measure the \ee\ hadronic cross section at \sqs\ smaller than the
initial \ee\ center-of-mass (CM) energy ($E_{CM}$) at $B$-factories,
ISR can be used~\cite{babar2}.  ISR allows a measurement of cross
sections in a broad energy range while the high luminosity of the
$B$-factories compensates for the suppression associated with the
emission of a hard photon. The selection of \eeddchg\ signal events
using full reconstruction of both the $D^{(*)+}$ and $D^{*-}$ mesons,
plus the \gisr\ photon, suffers from low efficiency due to the low
$D^{(*)}$ reconstruction efficiencies, small branching fractions and
the low geometrical acceptance for the \gisr, which tends to be
emitted along the beam line. Here, we use a method that achieves higher
efficiency by requiring full reconstruction of only one of the
$D^{(*)+}$ mesons, the \gisr, and the slow $\pi_{\mathrm{slow}}^-$
from the other $D^{*-}$. In this case the spectrum of masses recoiling
against the $D^{(*)+} \gisr$ system:
\begin{eqnarray}
\RM(D^{(*)+}\gisr)\!=\!\sqrt{(E_{CM}\!-\!E{}_{D^{(*)+} \gisr}){}^2\!-\!
p{}^2_{D^{(*)+} \gisr}}
\end{eqnarray}
peaks at the $D^{*-}$ mass. Here $E_{D^{(*)+} \gisr}$ and $p_{D^{(*)+}
\gisr}$ are the CM energy and momentum, respectively, of the $D^{(*)+}
\gisr$ combination. This peak is expected to be wide and asymmetric
due to the photon energy resolution and higher-order corrections to
ISR ({\it i.e.},  more than one \gisr\ in the event). From the Monte
Carlo (MC) simulation the resolution of this peak is estimated to be
$\sim 300\mevc$, which is not sufficient to separate the
$D\overline{D}{}^*$, $D^*\overline{D}{}^*$ or $\dd\pi$ final
states. To disentangle the contributions from these final states and
to suppress combinatorial backgrounds, we use the slow pion from the
unreconstructed $D^{*-}$. The difference between the mass recoiling
against $D^{(*)+} \gisr$ and $D^{(*)+} \pi_{\mathrm{slow}}^- \gisr$
(recoil mass difference):
\begin{eqnarray}
\Delta\RM \!= \!\RM(D^{(*)+}\gisr)\!-\!\RM(D^{(*)+}
\pi_{\mathrm{slow}}^- \gisr)\, ,
\end{eqnarray}
has a narrow distribution ($\sigma\!\sim\!1.4\mevc$) around the
nominal $m_{D^{*-}}- m_{\overline{D}{}^0}$, since the uncertainty in
\gisr\ momentum partially cancels out.

For the measurement of the exclusive cross section, one needs to
determine the \ddch\ mass when one of the $D^*$'s is not
reconstructed.  In the absence of higher-order QED processes,
$M(\ddch)$ is the mass recoiling against the \gisr. However, the
photon energy resolution results in a typical $\RM(\gisr)$ resolution
of $\sim 100 \mev$, which is too wide for the study of relatively
narrow \ddch\ mass states. We significantly improve the $\RM(\gisr)$
resolution by applying a refit that constrains $\RM(D^{(*)+} \gisr)$
to the nominal $D^{*-}$ mass. In this way we use the well-measured
properties of the fully reconstructed $D^{(*)+}$ to correct the energy
of the \gisr.  As a result, the $M_{\dd}$ ($ \equiv \RM(\gisr))$
resolution is improved by a factor of $\sim 10$ and varies from $\sim
6\mevc$ around threshold to $\sim 12 \mevc$ at $M_{D^{(*)+}D^{*-}}=5.0
\gevc$.  The recoil mass difference after the refit procedure
($\Delta\RMF$) has a resolution improved by a factor of $\sim
2$. Finally, the cross section is derived from the \ddch\ mass
spectrum after the refit.

All charged tracks should originate from the interaction point (IP)
with the requirements $dr<2 \, {\mathrm{cm}}$ and
$dz<4\,{\mathrm{cm}}$, where $dr$ and $dz$ are the impact parameters
perpendicular to and along the beam direction with respect to the
IP. Charged kaons are required to have the ratio of particle
identification likelihoods, $\mathcal{P}_K = \mathcal{L}_K /
(\mathcal{L}_K + \mathcal{L}_\pi)$~\cite{nim}, larger than 0.6.  No
identification requirements are applied for pion candidates. $K^0_S$
candidates are reconstructed by combining $\pi^+ \pi^-$ pairs with an
invariant mass within $10\mevc$ of the nominal $K^0_S$ mass. The
distance between the two pion tracks at the $K^0_S$ vertex must be
less than $1\,\mathrm{cm}$, the transverse flight distance from the
interaction point is required to be greater than $0.1\,\mathrm{cm}$,
and the angle between the $K^0_S$ momentum direction and the flight
direction in the $x-y$ plane should be smaller than
$0.1\,\mathrm{rad}$. Photons are reconstructed in the electromagnetic
calorimeter as showers with energies greater than $50 \mev$ that are
not associated with charged tracks. ISR photon candidates are required
to have energies greater than $2.5 \gev$. Pairs of photons are
combined to form $\pi^0$ candidates. If the mass of a $\gamma \gamma$
pair lies within $15\mevc$ of the nominal $\pi^0$ mass, the pair is
fit with a $\pi^0$ mass constraint and considered as a $\pi^0$
candidate.

$D^0$ candidates are reconstructed using five decay modes: $K^-
\pi^+$, $K^- K^+$, $K^- \pi^- \pi^+ \pi^+$, $K^0_S \pi^+\pi^-$ and
$K^- \pi^+ \pi^0$.  A $\pm 15\mevc$ mass window is used for all modes
except for $K^- \pi^- \pi^+ \pi^+$, where a $\pm 10\mevc$ requirement
is applied ($\sim 2.5\,\sigma$ in each case). $D^+$ candidates are
reconstructed using the decay modes $K^0_S\pi^+$, $K^- \pi^+ \pi^+$
and $K^- K^+ \pi^+$. A $\pm 10\mevc$ mass window is used for all $D^+$
modes. To improve the momentum resolution of $D$ meson candidates,
final tracks are fitted to a common vertex applying the nominal $D^0$
or $D^+$ mass as a constraint. $D^*$ candidates are selected via
$D^{*+} \to D^0 \pi^+$ and $D^{*0} \to D^0 \pi^0$ decay modes with a
$\pm 2\mevc$ $D^*-D$ mass-difference window.  A mass- and vertex-constrained
fit is also applied to $D^*$ candidates.

\begin{figure}[htb]
\begin{tabular}{cc}
\hspace*{-0.025\textwidth}
\includegraphics[width=0.48\textwidth]{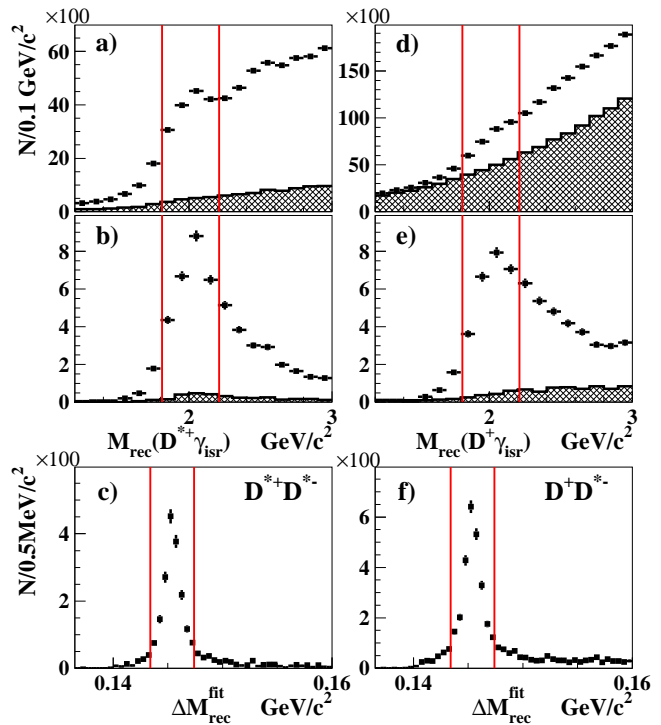}
\end{tabular}
\caption{The $\RM (D^{(*)+} \gisr)$ distribution for the data: a), d)
without requirement of slow pion; b), e) with $\Delta\RMF$
requirement. Histograms show the normalized $M_{D^{(*)+}}$ sidebands
contributions.  c), f) The distribution of $\Delta\RMF$ in the data
after the refit procedure. The selected signal windows are indicated
by the vertical lines.}
\label{Fig1}
\end{figure}

The distribution of $\RM (D^{*+} \gisr)$ in the data, without any
requirements on the slow pion, is shown in Fig.~\ref{Fig1}\,a). The
excess around the $D^{*-}$ mass includes the \eedpdmg\ signal as well
as contributions from the \eedpdstmg\ channel. The shoulder at higher
masses is due to $e^+ e^- \to D^{*+} D^{(*)} \pi \gisr$. The excess
that is evident at $\sim 2.5\gevc$ corresponds to $e^+ e^- \to D^{*+}
D^{**-} \gisr$. The background from the other processes is
substantially suppressed by the inclusion of the slow pion from the
unreconstructed $D^{*-}$ and the tight requirement on $\Delta\RMF$;
{\it i.e.}, within $\pm 2\mevc$ of the nominal $m_{D^{*-}}-
m_{\overline{D}{}^0}$, a clean peak corresponding to \eedpdm\ is
evident in the $\RM (D^{*+} \gisr)$ distribution
(Fig.~\ref{Fig1}\,b)).
We define the signal region by the requirement that $\RM(D^{*+}
\gisr)$ be within $\pm 0.2\gevc$ of the nominal $D^{*-}$ mass.  This
tight requirement suppresses $e^+ e^- \to D^{*+} D^{*-} \pi^0 \gisr$
events, which have a similar $\Delta\RMF$ distribution. The spectrum
of $\Delta\RMF$ for the signal $\RM(D^{*+} \gisr)$ window after the
refit procedure in data is shown in Fig.~\ref{Fig1}\,c).

In case of multiple entries, the $D^{*+}\pi$ candidate with the
minimum value of $\chi^2_{\mathrm{tot}} = \chi^2_{M(D^0)} +
\chi^2_{M(D^{*+})} + \chi^2_{\Delta\RMF}$ is chosen, where each
$\chi^2$ is defined as the squared ratio of the deviation of the
measured parameter from the expected signal value and the
corresponding resolution.

The following sources of background are considered: (1) combinatorial
background under the reconstructed $D^{*+}$ peak; (2) real $D^{*+}$
mesons coming from the signal or other processes combined with a
combinatorial slow pion; (3) both the $D^{*+}$ and slow pion are
combinatorial; (4) the reflection from the process \dpdmpi\ with an
extra $\pi^0$ in the final state; (5) the contribution of \dpdmis\
when an energetic $\pi^0$ is misidentified as a single \gisr.  The
contributions of backgrounds (1) and (2) are extracted using $D^{*+}$
and $\Delta\RMF$ sidebands, which have twice the area of the signal
region. The latter sideband is shifted by $10\mevc$ from the signal
region to avoid signal over-subtraction due to the higher-order ISR
tail.  Background (3) is present in both the $M(D^{*+})$ and
$\Delta\RMF$ sidebands and is, thus, subtracted twice. To account for
this over-subtraction we use a 2-dimensional sideband region, where
events are selected from both the $D^{*+}$ and the $\Delta\RMF$
sidebands. The total fraction of combinatorial backgrounds (1--3),
found to be as large as $\sim 10\%$, is subtracted from the
signal-region \dpdm\ mass spectrum.

Process (4) produces a broad peak in the $\Delta\RMF$ distribution
around the nominal $m_{D^{*-}}- m_{\overline{D}{}^0}$ value and, thus,
is not contained in the $\Delta\RMF$ sidebands. The dominant part of
this background is suppressed by the tight requirement on $\RM(D^{*+}
\gisr)$.  The remaining part is estimated by applying a similar
partial reconstruction method to the isospin-conjugate process $e^+
e^- \to D^{*0} D^{*-} \pi^+_{\mathrm{miss}}\gisr$: a $D^{*0}$ is fully
reconstructed and the slow pion from a $D^{*-}$ is used for the
$\Delta\RMF$ requirement. Since there is a charge imbalance in this
final state, only events with a missing extra $\pi^+_{\mathrm{miss}}$
can contribute to the $\Delta\RMF$ peak. To extract the level of
background (4), this spectrum is rescaled according to the ratio of
$D^{*0}$ and $D^{*+}$ reconstruction efficiencies and an isospin
factor of 1/2. The contribution from background (4) is found to be
consistent with zero. Uncertainties in this estimate are included in
the systematic error. The contribution of background (5) is determined
from reconstructed \dpdmis\ events using a similar partial
reconstruction technique but with an energetic $\pi^0$ replacing the
\gisr. The contribution of this background is found to be negligibly
small; uncertainties in this estimate are also included in the
systematic error.

The analysis of the \eedpdstm\ cross section is identical to that
described above for \eedpdm\ with the fully reconstructed $D^{*+}$
meson replaced by a fully reconstructed $D^+$ meson. The distribution
of $\RM (D^+\gisr)$ with no requirements on the slow pion from the
$D^{*-}$ is shown in Fig.~\ref{Fig1}\,d). The excess around the
nominal $D^{*-}$ mass corresponds to the \eedpdstmg\ signal plus
overlaps from the \eedpdmg\ channel. The shoulder at higher masses is
due to $e^+ e^- \to D^{(*)+} D^{*-} \pi \gisr$. The requirement of a
detected slow pion from the unreconstructed $D^{*-}$ and a tight
requirement on $\Delta\RMF$ provides the clean \eedpdstm\ signal peak
in the distribution of $\RM (D^+\gisr)$ that is shown in
Fig.~\ref{Fig1}\,e).  The $\Delta\RMF$ distribution for the signal
$\RM(D^+\gisr)$ window is shown in Fig.~\ref{Fig1}\,f).  In the case
of multiple entries, we apply a single candidate selection procedure
similar to that used for \dpdm\ to all the distributions shown below.

Similar sources of background (1--5) are considered for the \eedpdstm\
study. The contributions of backgrounds (1--3) are determined using
$D^+$ and $\Delta\RMF$ sidebands with twice the area of the signal
region. Here again, background (3) is present in both $M(D^+)$ and
$\Delta\RMF$ sidebands and thus subtracted twice. To account for this
over-subtraction we use a 2-dimensional sideband region that contains
pure background (3) events.

The level of contamination from background (4) is determined from
isospin-conjugate events, \dndstg, in the data. The \dpdstm\ analysis
is repeated with the fully reconstructed $D^+$'s replaced by fully
reconstructed $D^0$'s. The $D^0 D^{*-}$ mass distribution, after
combinatorial background subtraction, is rescaled according to the
ratio of $D^+$ and $D^0$ reconstruction efficiencies and an isospin
factor of 1/2. By chance, the contamination from the process $e^+ e^-
\to D^{*+} D^{*-}$, followed by $D^{*+} \to D^+ \pi^0$ is also
included in our estimate with correct scaling because
$\mathcal{B}(D^{*+}\to D^+ \pi^0) / \mathcal{B}(D^{*+} \to D^0 \pi^+)$
is also approximately 1/2.  The contribution from background (5)
determined from reconstructed \dpdstmis\ events in the data is found
to be negligibly small and taken into account in the systematic error.
The total background level is $\sim20\%$ or less of the signal for all
values of $M(D^+D^{*-})$.

\begin{figure}[htb]
\hspace*{-0.025\textwidth}
\includegraphics[width=0.48\textwidth]{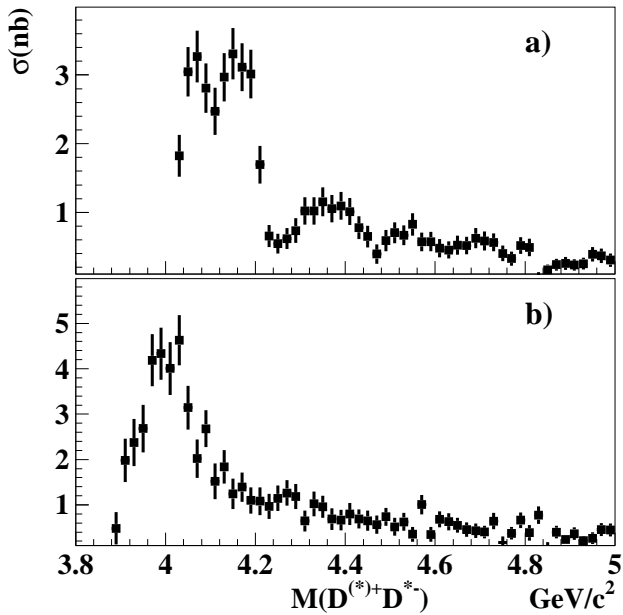}
\caption{The exclusive cross sections for a) \eedpdm\ and b) \eedpdstm
. The error bars include only the statistical uncertainties.}
\label{Fig2}
\end{figure}

The \eeddch\ cross sections are extracted from the \ddch\ mass
distributions by the relation~\cite{babar3}
\begin{eqnarray}
\sigma(e^+e^- \to D^{(*)+} D^{*-}) = \frac{ dN/dm }{ \eta_{\mathrm{tot}} dL/dm} \, ,
\end{eqnarray}
where $m\equiv M(\ddch)$, $dN/dm$ is the mass spectra obtained before
corrections for resolution and higher-order radiation, while
$\eta_{\mathrm{tot}}$ is the total efficiency. The factor $dL/dm$ is
the differential ISR luminosity
\begin{eqnarray}
 dL/dm =\frac{\alpha}{\pi x}\Bigl((2-2x+x^2)\ln\frac{1+C}{1-C}-x^2C\Bigr)
\frac{2m \mathcal{L}}{E^2_{\mathrm{CM}}} \, ,
\end{eqnarray}
where $x = 1 - m^2/E^2_{\mathrm{CM}}$, $\mathcal{L}$ is the total
integrated luminosity and $C = \cos\theta_0$, where $\theta_0$ defines
the polar angle range for \gisr\ in the \ee\ CM frame:
$\theta_0<\theta_{\gisr}<180-\theta_0$. The reconstruction
efficiencies, determined as a function of $M(\ddch)$ by MC simulation,
are found to be independent of $M(\ddch)$ for both processes and are
equal to $\eta(\dpdm) = 4.3\times 10^{-3}$ and $\eta(\dpdstm) = 3.9
\times 10^{-3}$. The resulting exclusive \eeddch\ cross sections are
shown in Fig.~\ref{Fig2} with statistical uncertainties only.
Since the bin width is much larger than the resolution, no correction
for resolution is applied.  Since a reliable fit to the cross-sections
obtained above requires a solution to a non-trivial and
model-dependent problem of coupled channels and threshold effects, we
do not report results here.

The systematic errors for the $\sigma$(\eeddch) measurements are
summarized in Table~\ref{sys1}.
\begin{table}[htb]
\caption{Contributions to the systematic error on the cross sections,
[$\%$].}
\label{sys1}
\begin{center}
\begin{tabular}{@{\hspace{0.4cm}}l@{\hspace{0.4cm}}||@{\hspace{0.4cm}}c@{\hspace{0.4cm}}|@{\hspace{0.5cm}}c@{\hspace{0.4cm}}}
\hline \hline
Source                     & \dpdm\   & $D^+ D^{*-}$ \\
\hline
Background subtraction     & $\pm 5$  &  $\pm 4$     \\
Angular distributions      & $\pm 5$  &  ---         \\
Reconstruction             & $\pm 7$  &  $\pm 6$     \\
Cross section calculation  & $\pm 2$  &  $\pm 2$     \\
$\mathcal{B}(D^{(*)})$     & $\pm 3$  &  $\pm 5$     \\
MC statistics              & $\pm 4$  &  $\pm 3$     \\
Kaon identification        & $\pm 1$  &  $\pm 1$     \\
\hline
Total                      & $\pm 11$ &  $\pm 10$     \\
\hline \hline
\end{tabular}
\end{center}
\end{table}
The systematic errors associated with the background (1--3)
subtraction are estimated to be 3\% from the uncertainty in the
scaling factors for the sideband subtractions using fits to the
$M(D^{(*)+})$ and $\Delta \RMF$ distributions in the data with
different signal and background parametrization.  Uncertainties in
backgrounds (4--5) are estimated conservatively to be smaller than 2\%
of the signal in the case of \dpdm; these two sources are added
linearly to give 5\% in total. In the case of the \dpdstm, backgrounds
(4-5) are subtracted using the data and only the uncertainty in the
scaling factor for the subtracted distribution is taken into account.
A second source of systematic error is due to the unknown helicity
angle composition of the \dpdm\ final state which can be a mixture of
$D^{*+}_T D^{*-}_T$ $D^{*+}_T D^{*-}_L$ and $D^{*+}_L D^{*-}_L$, where
the subscripts $L$ and $T$ refer to longitudinally and transversely
polarized $D^*$'s. For the efficiency calculation, we assume equal
fractions of these helicity states and consider the extreme cases
(pure $D^{*+}_T D^{*-}_T$ and pure $D^{*+}_L D^{*-}_L$) for the
efficiency uncertainty. There is no corresponding uncertainty in the
case of the $D^{+}D^{*-}$ final state, where the $D^{*-}$ polarization
is fixed by angular momentum and parity conservation.  A third source
of systematic error comes from the uncertainties in track and photon
reconstruction efficiencies, which are 1\% per track, 2\% per slow
pion and 1.5\% per photon, respectively. The systematic error ascribed
to the cross section calculation is estimated from a study of the
$C$-dependence of the final result and includes a 1.5\% error on the
total luminosity. Other contributions come from the uncertainty in the
identification efficiency and the absolute $D^0$ and $D^{(*)+}$
branching fractions~\cite{pdg}.

In summary, we report the first measurements of exclusive \eedpdm\ and
\eedpdstm\ cross sections at \sqs\ around the \dpdm\ and \dpdstm\
thresholds with initial state radiation.  The shape of the \eedpdm\
cross section is complicated with several local maxima and minima. The
minimum near 4.25\gevc\,---\,in the Y(4260) region\,---\,could be due
to $D_s^* D_s^*$ ($D D^{**}$) threshold effects described
in~\cite{voloshin1,voloshin2,rosner} or due to destructive
interference of this state with other $\psi(nS)$ states.  Aside from a
prominent excess near the $\psi(4040)$, the \eedpdstm\ cross section
is relatively featureless. The measured cross sections are
compatible~\cite{foot2} within errors with the $D^{(*)}\overline
{D}{}^{*}$ exclusive cross section in the energy region up to
$4.260\gev$ measured by CLEO-c~\cite{cleo1}.

We thank the KEKB group for excellent operation of the accelerator,
the KEK cryogenics group for efficient solenoid operations, and the
KEK computer group and the NII for valuable computing and Super-SINET
network support.  We acknowledge support from MEXT and JSPS (Japan);
ARC and DEST (Australia); NSFC and KIP of CAS (China); DST (India);
MOEHRD, KOSEF and KRF (Korea); KBN (Poland); MIST (Russia); ARRS
(Slovenia); SNSF (Switzerland); NSC and MOE (Taiwan); and DOE (USA).


\begin{thebibliography} {99}
\bibitem{foot1} Charge-conjugate modes are included throughout this
paper.
\bibitem{foot2} Since only charged final states are measured, our
results should be scaled by a factor of two for this comparison.
\end{thebibliography}
\end{document}